\documentclass[secnumarabic,amssymb,showpacs, nobibnotes, aps,showkeys, prd]{revtex4}
\usepackage{graphics}
\usepackage{color}
\usepackage{amssymb}
\usepackage{latexsym}
\usepackage{longtable}
\begin{document}
\keywords{Entropic Gravity, Axial Anomaly.}
\pacs{04.20.-q, 04.20.Cv}
\title{\textit{Axial Current, Killing Vector and  Newtonian Gravity}}
\author{Prasanta Mahato\thanks{E-mail: pmahato@dataone.in}}
\affiliation{Narasinha Dutt College,  Howrah, West Bengal, India 711 101}
\begin{abstract}
  Starting from the multiplicative torsion approach of gravity and assuming a Killing vector to be  proportional to the axial-vector matter current, here we derive Newton's law of gravity where the logarithm of the proportionality factor has been found to be the potential function.
\end{abstract}
\maketitle  
\section{From Axial Current to Newtonian Gravity}

All we know that the fermions as main building blocks of
matter are  described by spinor fields. In contrast,
the interactions are mediated by bosons. Any realistic
spinor theory has therefore to account for the bosons as to be composed of an even number of fermions. It is generally beleived that there is a  Fermi sea of Neutrinos in the universe and the magnitude of the Fermi energy $E_F$ is related to various cosmological theories\cite{Wei62}. Hence it is quite plausible to consider that spinors, massive or massless, are omnipresent at each space-time point of the universe.

 By Geroch's theorem\cite{Ger68} we know that - the existence of  the spinor structure is equivalent to the existence of a global field of orthonormal tetrads on the space and time orientable manifold. This requires spinors to be soldered with the tetrads at each space-time point.
The idea of gravity emerging from spinors is not new
and fairly obvious, as one can construct a spin-2 particle
as the direct product of spinors\cite{Kra,Oha}.  The first idea of
this type is due to Bjorken\cite{Bjo}, who attempted to formulate
the photon and graviton as a composite state. Another  successful attempt is due to
Hebecker and Wetterich\cite{Wet}. Their theory can be regarded
as a reformulation of gravity in terms of spinors.

In multiplicative torsion approach of gravity\cite{MPLAPRD,Mah05,Mah07a,Mah07b,Mah09,Mah09a,Mah11} one gets axial vector current 1-form for spinor field $\Psi$, $J_5=\bar{\Psi}\gamma_5\gamma\Psi$, to be an exact 1-form, given by the equation
\begin{eqnarray}
d(\mathcal{R}-\beta\phi^2-\frac{1}{2}\kappa m_\Psi\bar{\Psi}\Psi)=-\frac{g}{4}\bar{\Psi}\gamma_5\gamma\Psi,\label{eqn:ng001}\\\mbox{where,}\hspace{1mm}\gamma\equiv\gamma_\mu dx^\mu=\gamma_a e^a{}_\mu dx^\mu\equiv\gamma_a e^a,\nonumber
\end{eqnarray} here  Latin  and Greek indices signify  local tangent space and external coordinates of the four dimensional space-time manifold, respectively. Here the local flat-space metric is given by $
\eta_{ab}=\left
(\begin{array}{rrrr}-1&0&0&0\\0&1&0&0\\0&0&1&0\\0&0&0&1\end{array}\right)$. $\mathcal{R}$ is the Ricci scalar and $\phi$ is a variable
parameter of dimension $(length)^{-1}$ and Weyl weight $(-1)$, such
that $\phi e^a$, where $e^a$ is the local frame field, has the correct  dimension and conformal weight
of the de Sitter boost part of the $SO(4,1)$ gauge connection\cite{Mah05,Mah07a}. $\phi$ may be linked with the dark matter where $\sqrt{\frac{\beta}{\kappa}}$ is the  mass of the dark field\cite{Mah07a}. Recently it has been shown that, from $J_5$ it is possible to construct an axial vector current 3-form which is  conserved according to  Noether's theorem\cite{Mah09}.  If we neglect mass terms, then (\ref{eqn:ng001}) reduces to 
\begin{eqnarray}
d\mathcal{R}=-\frac{g}{4}J_5\label{eqn:ng002}
\end{eqnarray}
We know that any  Killing vector  $\xi$ satisfies\cite{Pad10} 
\begin{eqnarray}
\xi\wedge{}^*d\mathcal{R}=0\label{eqn:ng003}
\end{eqnarray}From (\ref{eqn:ng002}) \& (\ref{eqn:ng003}) we can write 
\begin{eqnarray}
\xi\wedge{}^*J_5=0\label{eqn:ng004}
\end{eqnarray}In absence of mass, $J_5$ may be assumed to be a vector on the light cone. Then (\ref{eqn:ng004}) implies that $\xi\propto J_5$ or,
\begin{eqnarray}
\xi=\chi J_5, \label{eqn:ng005}
\end{eqnarray}where $\chi$ is a dimensional scalar factor. Here we   define the surface gravity $\kappa_\xi$ for the Killing vector $\xi$ given by 
\begin{eqnarray}
\xi^\mu\bar{\nabla}_\mu\xi^\nu=\kappa_\xi\xi^\nu\nonumber\\\mbox{or,}\hspace{1mm}\xi\wedge{}^*d\xi=2\kappa_\xi{}^*\xi, \label{eqn:ng006}
\end{eqnarray}where $\bar{\nabla}$ is the torsion-free covariant derivative.  Now  $\xi$ being  a null vector, after using (\ref{eqn:ng005}) \& (\ref{eqn:ng006}), we  can write
\begin{eqnarray}
\kappa_\xi=-\xi^\mu\bar{\nabla}_\mu V\nonumber\\\mbox{where,}\hspace{2mm}V=-\frac{1}{2}\ln \chi\label{eqn:ng007}
\end{eqnarray}Here we like to consider   (\ref{eqn:ng005}) \& (\ref{eqn:ng007}) as \textbf{two postulates} to be true for any  axial current $J_5$ of arbitrary spinors (massive or massless) in any space-time point.
Using exactness of $J_5$ and taking exterior derivatives of (\ref{eqn:ng005}), we get 
\begin{eqnarray} 
   d\xi &=&-2dV\wedge\xi,\hspace{2mm}\label{eqn:ng009}\\
   \mbox{or,}\hspace{2mm}d{}^*(d\xi)&=&-2d{}^*(dV\wedge\xi)\nonumber\\\mbox{or,}\hspace{2mm}
   -\bar{R}_{\mu\nu}\xi^{\nu}&=&-\square V\xi_\mu+\partial_\mu\kappa_\xi,\label{eqn:ng010}   
\end{eqnarray}here $\square V\equiv\bar{\nabla}_a\bar{\nabla}^a V$ and $\bar{R}_{\mu\nu}$ is the torsion-free Ricci tensor.

Using  properties of Killing vectors, we may write,
\begin{eqnarray}
0=d{}^*\xi=d\chi\wedge{}^*J_5+\chi d{}^*J_5\nonumber\\\mbox{or,}\hspace{1mm}\chi d{}^*J_5=-2\xi\wedge{}^*dV=2\kappa_\xi\eta,\label{eqn:ng008}
\end{eqnarray}here $\eta(=\frac{1}{4!}e^a\wedge e^b\wedge e^c\wedge e^d)$ is the invariant volume 4-form. This last equation implies that axial anomaly is proportional to the surface gravity $\kappa_\xi$. We know that, in  theories like superstring theories and lattice gauge theories, anomaly  cancellation takes place\cite{Anomaly}.  Again we know that $\kappa_\xi$ must be a constant when $\xi$ is along a null geodesic
generator of any Killing horizon\cite{Wal99}. So from last equation, for the time being, we may assume $\kappa_\xi=constant$ and putting it  in (\ref{eqn:ng010})  we get, w.r.t. local indices,
\begin{eqnarray}
\square V\xi^a= \bar{R}^{a}{}_{b}\xi^{b}\label{en:ng011}
\end{eqnarray}This equation implies that $\square V$ is an eigen value of the matrix $\bar{R}^{a}{}_{b}$ with $\xi^{b}$ being the corresponding eigen vector in the  local Minkowski space. 

For massive spinors, we may consider,  the vector current $J$ to be time-like and the axial vector current $J_5$ to be space-like. Here we may compare the pair ($J$,$J_5$)  with the pair ($p$,$s$)  where $p$ is the timelike momentum vector and  $s$ is the spacelike spin vector of an one particle state, s.t. $p^as_a=0$. This implies that  $\xi$
is also a space-like killing vector and is asociated with the spin degree of freedom of matter\cite{Mie06}.

\begin{enumerate}
	\item \underline {At first we consider the case where the killing vector is space-like.}
	
\begin{enumerate}
	\item For a region dominated by a pressureless isotropic matter with rest mass density $\rho$, we may take, 

\begin{eqnarray}
\xi^a&=&\vartheta(0,l,m,n),\hspace{1mm}l^2+m^2+n{}^2=1\hspace{1mm}\mbox{and}\hspace{1mm}\vartheta\hspace{1mm}\mbox{is a nonzero scalar},\label{eqn0002}\\\bar{R}^{a}{}_{b}&=&4\pi G\rho\left
(\begin{array}{rrrr}-1&0&0&0\\0&1&0&0\\0&0&1&0\\0&0&0&1\end{array}\right),\hspace{2mm}\mbox{$G$ is Newton's constant,}\\\bar{G}^{a}{}_{b}&=&4\pi G\rho\left
(\begin{array}{rccc}-2&0&0&0\\0&0&0&0\\0&0&0&0\\0&0&0&0\end{array}\right).
\end{eqnarray}With this form of the matrix $\bar{R}^{a}{}_{b}$, (\ref{en:ng011}) reduces to 
\begin{eqnarray}
\square V=4\pi G\rho\label{en:ng016}
\end{eqnarray}Equations (\ref{en:ng016}) may be interpreted  as the generalisation of  the Poisson's equation of Newtonin gravity in curved space-time. In the case of weak field  approximation of a static mass distribution  this equation reduces to  Poisson's, given by 
\begin{eqnarray}
\bar{\nabla}_\mu\bar{\nabla}^\mu V\thickapprox{\mathop \nabla\limits^ \to }.{\mathop \nabla\limits^ \to }V=4\pi G\rho\label{en:ng017}\\\mbox{where,}\hspace{1mm}{\mathop \nabla\limits^ \to }.{\mathop \nabla\limits^ \to }\equiv\nabla^2=\frac{\partial^2}{\partial x^2}+\frac{\partial^2}{\partial y^2}+\frac{\partial^2}{\partial z^2}.\nonumber
\end{eqnarray}

\item For a region dominated by  a radiation fluid with energy density $\rho$, we may take,

\begin{eqnarray}
\bar{R}^{a}{}_{b}=\bar{G}^{a}{}_{b}=8\pi G\rho\left
(\begin{array}{rrrr}-1&0&0&0\\0&l^2&lm&ln\\0&lm&m^2&mn\\0&ln&mn&n^2\end{array}\right)\label{eqn321}
\end{eqnarray}With this form of the matrix $\bar{R}^{a}{}_{b}$, (\ref{en:ng011}) reduces to 
\begin{eqnarray}
\square V=8\pi G\rho\label{en:ng0016}
\end{eqnarray}Equations (\ref{en:ng016}) and (\ref{en:ng0016}) justify that
radiation produces a gravitational potential which is
twice as strong as that of a material particle with the same energy density\cite{Pad10}.
\end{enumerate} 
\item \underline {Now we consider the case  where the killing vector is a null  vector.}
\begin{enumerate}
\item
For a region dominated by pressureless isotropic matter with rest mass density $\rho$, we may take,\begin{eqnarray}
\xi^a=\vartheta(1,l,m,n).\label{eqn0001}
\end{eqnarray}
\begin{eqnarray}
\bar{R}^{a}{}_{b}&=&4\pi G\rho\left
(\begin{array}{rrrr}-1&l&m&n\\-l&1&0&0\\-m&0&1&0\\-n&0&0&1\end{array}\right),
\\\bar{G}^{a}{}_{b}&=&4\pi G\rho\left
(\begin{array}{rrrr}-2&l&m&n\\-l&0&0&0\\-m&0&0&0\\-n&0&0&0\end{array}\right).
\end{eqnarray}
With this form of the matrix $\bar{R}^{a}{}_{b}$, (\ref{en:ng011}) reduces to 
\begin{eqnarray}
\square V=0\label{en:ng000167}
\end{eqnarray}
\item For a region dominated by  isotropic radiation fluid with energy density $\rho$, we may take,

In this case we may take
\begin{eqnarray}
\bar{R}^{a}{}_{b}=\bar{G}^{a}{}_{b}&=&8\pi G\rho\left
(\begin{array}{rrrr}-1&l&m&n\\-l&l^2&lm&ln\\-m&lm&m^2&mn\\-n&ln&mn&n^2\end{array}\right).\label{eqn:ng0015a}\end{eqnarray}

With this form of the matrix $\bar{R}^{a}{}_{b}$, (\ref{en:ng011}) reduces to equation (\ref{en:ng000167}). Here we see that, for $\xi$ being a null vector, the forms of $\bar{R}^{a}{}_{b}$ fail to produce the standard results for $V$ to be a gravitational potential of Newtonian gravity, i.e. equations (\ref{en:ng016}) and (\ref{en:ng0016}).
\end{enumerate}
\end{enumerate}
In view of space isotropy, we may consider $\bar{G}^{a}{}_{b}$ to be of the general form
\begin{eqnarray}\bar{G}^{a}{}_{b}&=&F\left
(\begin{array}{rccc}-1&dl&dm&dn\\-dl&bl^2+c&blm&bln\\-dm&blm&bm^2+c&bmn\\-dn&bln&bmn&bn^2+c\end{array}\right);\hspace{1mm}b,c,d\hspace{1mm}\mbox{are scalars and}\hspace{1mm}F=8\pi G\rho\label{en:ng012}
\end{eqnarray}   
Then
\begin{eqnarray}
\mathcal{R}&=&F(1-b-3c),\label{eqn:ng013}\\\bar{R}^{a}{}_{b}&=&F\left
(\begin{array}{cccc}\frac{-1-b-3c}{2}&dl&dm&dn\\-dl&bl^2+\frac{1-b-c}{2}&blm&bln\\-dm&blm&bm^2+\frac{1-b-c}{2}&bmn\\-dn&bln&bmn&bn^2+\frac{1-b-c}{2}\end{array}\right).\label{eqn:ng014}
\end{eqnarray}

If we take average over the two dimensional unit sphere $l^2+m^2+n^2=1$ then (\ref{en:ng012}) reduces to its isotropic form, given by
\begin{eqnarray}<\bar{G}^{a}{}_{b}>&=&F\left
(\begin{array}{rccc}-1&0&0&0\\0&\frac{b}{3}+c&0&0\\0&0&\frac{b}{3}+c&0\\0&0&0&\frac{b}{3}+c\end{array}\right),\mbox{where}\int l=\int m=\int n=0,\nonumber\\&{}&\int l^2=\int m^2=\int n^2=\frac{4\pi}{3}\hspace{1mm}\mbox{and}\hspace{1mm}\int lm=\int ln=\int mn=0.\label{en:ng00112}
\end{eqnarray} This form of $<\bar{G}^{a}{}_{b}>$ represents  an ideal radiation fluid, ideal pressureless fluid or dark energy according as $b+3c=1$, $0$ or $-3$.

Now we consider the solution of equation (\ref{en:ng011}) in the general case where $\mathcal{R}=-2AF$, s.t. $b+3c=1+2A$,
\begin{enumerate}
	\item 

With $\xi^a=\vartheta(1,l,m,n)$, we have the following solution
\begin{eqnarray}
\bar{G}^{a}{}_{b}&=&F\left
(\begin{array}{cccc}-1&(1+A+\theta)l&(1+A+\theta)m&(1+A+\theta)n\\-(1+A+\theta)l&(1+2A+3\theta) l^2-\theta&(1+2A+3\theta) lm&(1+2A+3\theta) ln\\-(1+A+\theta)m&(1+2A+3\theta) lm&(1+2A+3\theta)m^2-\theta&(1+2A+3\theta) mn\\-(1+A+\theta)n&(1+2A+3\theta) ln&(1+2A+3\theta) mn&(1+2A+3\theta) n^2-\theta\end{array}\right),\nonumber\\
{}&=&\bar{R}^{a}{}_{b}+AF\delta^a{}_b\label{eqn:ng01013115a}\\\mbox{where},&{}&\frac{b-1-2A}{3}=-c=d-1-A=\theta;\nonumber\\\mbox{s.t.}&{}&\square V=8\pi G\rho\theta\hspace{1mm}\mbox{and}<\bar{G}^{a}{}_{b}>=8\pi G\rho\left
(\begin{array}{cccc}-1&0&0&0\\0&\omega&0&0\\0&0&\omega&0\\0&0&0&\omega\end{array}\right)\hspace{1mm}\mbox{where}\hspace{1mm}\omega=\frac{1+2A}{3}.\label{eqn:ng01012215a}\end{eqnarray}
\item 

With $\xi^a=\vartheta(0,l,m,n)$, we have the following solution
\begin{eqnarray}
\bar{G}^{a}{}_{b}&=&F\left
(\begin{array}{cccc}-1&0&0&0\\0&\frac{A+3\theta-1}{2} l^2+\frac{1+A-\theta}{2}&\frac{A+3\theta-1}{2} lm&\frac{A+3\theta-1}{2} ln\\0&\frac{A+3\theta-1}{2} lm&\frac{A+3\theta-1}{2}m^2+\frac{1+A-\theta}{2}&\frac{A+3\theta-1}{2} mn\\0&\frac{A+3\theta-1}{2} ln&\frac{A+3\theta-1}{2} mn&\frac{A+3\theta-1}{2} n^2+\frac{1+A-\theta}{2}\end{array}\right)\label{eqn:ng010131155a}\\{}&=&\bar{R}^{a}{}_{b}+AF\delta^a{}_b\\\mbox{where},&{}&\frac{2b+1-A}{3}=1+A-2c=\theta,d=0, \omega=\frac{1+2A}{3};\nonumber\\\mbox{s.t.}&{}&\square V=8\pi G\rho\theta\hspace{1mm}\mbox{and}<\bar{G}^{a}{}_{b}>=8\pi G\rho\left
(\begin{array}{cccc}-1&0&0&0\\0&\omega&0&0\\0&0&\omega&0\\0&0&0&\omega\end{array}\right)\nonumber\end{eqnarray}
\end{enumerate}
It is to be noted that role of $\theta$ is significant in the field equation of the potential function $V$ but $\theta$ disappears from  the isotropic average expression of the energy-momentum tensor. Hence isotropic form of the matter and $\theta$ are hitherto unrelated.

Considering standard results of General Relativity\cite{Pad10},  equation (\ref{eqn:ng01012215a})
 encompasses Newtonian Gravity for the following values of the parameters $\omega$
and $\theta$:
\begin{itemize}
	\item Radiation $\Rightarrow$ $\omega=\frac{1}{3}$, $\theta=1$;
	\item Mass $\Rightarrow$ $\omega=0$, $\theta=\frac{1}{2}$;
	\item Dark Energy $\Rightarrow$ $\omega=-1$, $\theta=-1$,
	 
\end{itemize}
These values of $\omega$
and $\theta$ give us the relation
\begin{eqnarray}
\theta=\frac{3\omega+1}{2}\{1+\omega(\omega-\frac{1}{3})(\omega+1)\epsilon(\omega)\},\label{eqn:ng00888}
\end{eqnarray}where $\epsilon(\omega)$ is  an arbitrary function of $\omega$. 
\begin{itemize}
	\item 
As a special case we consider $\epsilon(\omega)=0$, s.t.
 $\theta=\frac{3\omega+1}{2}$,  having $<T_{ab}>=$ diagonal $(\rho,p,p,p)$ and $p=\omega\rho$ is the equation of state of the isotropic matter. Using this relation in equation (\ref{eqn:ng01012215a}), we get the standard FRW result\cite{Pad10}
\begin{eqnarray}&{}&\square V=4\pi G\rho(1+3\omega)\hspace{1mm}\mbox{and}<\bar{G}^{a}{}_{b}>=8\pi G\rho\left
(\begin{array}{cccc}-1&0&0&0\\0&\omega &0&0\\0&0&\omega &0\\0&0&0&\omega \end{array}\right).\label{eqn:ng0110122155a}\end{eqnarray}It is well known that the curvature energy density corresponding to the spatial hypersurfaces of the Friedmann universe does not act as a source of gravitational potential. Here the case is  given by $\omega=-\frac{1}{3}$, i.e. $p=-\frac{1}{3}\rho$. In this case, for both null or spacelike $\xi^a$, 
\begin{eqnarray}
&{}&\bar{G}^{a}{}_{b}=\bar{R}^{a}{}_{b}-F\delta^a{}_b= F\left
(\begin{array}{cccc}-1&0&0&0\\0&-l^2&-lm&-ln\\0&-lm&-m^2&-mn\\0&-ln&-mn&-n^2\end{array}\right)\label{eqn:ng0510131155a}\\\mbox{s.t.}&{}&\square V=0\hspace{1mm}\mbox{and}<\bar{G}^{a}{}_{b}>=8\pi G\rho\left
(\begin{array}{rrrr}-1&0&0&0\\0&-\frac{1}{3}&0&0\\0&0&-\frac{1}{3}&0\\0&0&0&-\frac{1}{3}\end{array}\right).\label{eqn:ng0160122155a}\end{eqnarray}
\end{itemize}

\section{From Newtonian Gravity to Holographic Principle}
Recently, Verlinde\cite{Ver10} has proposed a remarkable
new idea  linking classical gravity to entropic force,
which attracted much interest\cite{Ent-Grav}. He has derived Newton's second law and  Einstein's equation from the relation between the entropy of a holographic screen and the mass inside the screen. Padmanabhan\cite{Padmanabhan}, earlier than Verlinde, has also proposed that
classical gravity can be derived from the equipartition energy of horizons. 
 Let us try to understand, in brief, this holographic nature of gravity.

Let, at time $t$ , P be a test particle  at a distance $r$ from a mass $m$ at O. Draw a spherical holographic screen through P having centre at O. Information of the mass $m$ takes a time $\Delta t=\frac{r}{c}$,  $c$ is the  velocity of light,  to reach the spherical boundary. Therefore at any time $t$ only  the whole of the holographic screen carries the net information of $m$  at O which originates from a past time $t-\Delta t$. Total number of bits available for carrying this information of mass $m$ at time $t$ is 
\begin{eqnarray}N=\frac{A}{l_P^2}=\frac{4\pi r^2c^3}{G\hbar}, \hspace{1mm}G=\mbox{Newton's Constant.}\label{en:ng018}\end{eqnarray} From definition of temperature, using equipartition rule\cite{Padmanabhan}, we have \begin{eqnarray}E=mc^2=\frac{1}{2}Nk_BT \label{en:ng019}\end{eqnarray} and then identifying $T$ with Unruh temperature\cite{Unr76} \begin{eqnarray}T=\frac{\hbar a}{2\pi k_Bc},\label{en:ng020}\end{eqnarray} we get Newton's  law for acceleration \begin{eqnarray}a=\frac{Gm}{r^2}.\label{en:ng021}\end{eqnarray} This derivation of Newton's law of gravitation is more than a `action at a distance' in nature. The holographic view emerges from the non-instataneous ability of signals [$\square V=0$ in equation (\ref{eqn:ng0160122155a})!], carrying information of the mass $m$, to reach the spherical boundary. This holographic  origin of gravity  claims it (gravity) to be an entropic force!

  We see that   (\ref{en:ng017}) is Poisson's equation for a static mass density $\rho$. This  equation also implies Newtonian gravity, i.e., acceleraton $a$ of a test particle, due to a point mass $m$  at a distance $r$, is given by (\ref{en:ng021}), where $\kappa =\frac{8\pi G}{c^2}$. 
  
  Starting from (\ref{en:ng021}), defining Unruh temperature by (\ref{en:ng020}) and then using thermodynamic relation  (\ref{en:ng019}) we get (\ref{en:ng018}) in the reverse order. Thus the holographic principle, i.e., the maximal storage space, or total number of bits, is proportional to the area $A$, is \textbf{a consequence of Newtonian form of gravity.} It is to be noted that in this approach the role of Newton's constant in the holographic principle \textbf{as the minimal unit of surface area is not by an ad-hoc prescription.} In multiplicative torsion approach of gravity the emergence of Newton's constant is through field equations\cite{MPLAPRD,Mah05,Mah07a}. $\kappa$ has topological origin, it is inversely proportional to the topological Nieh-Yan density.  It is to be noted that, in GR, one gets Newtonian Gravity only when one applies  weak field approximation on the metric's time component. But in the present formalism the metric has no such direct role. Even, after deriving equations (\ref{en:ng016}) \& (\ref{en:ng000167}), in the Minkowskian limit 
\begin{eqnarray}  
  g_{\mu\nu}\longrightarrow \left
(\begin{array}{cccc}1&0&0&0\\0&-1&0&0\\0&0&-1&0\\0&0&0&-1\end{array}\right),\nonumber
\end{eqnarray}these equations reduce to standard equations of special theory of relativity. Here (\ref{en:ng000167}) implies that, in a place having zero mass distribution, $V$ propagates with velocity of light and in static case, having some non-zero mass distribution, (\ref{en:ng016}) reduces to Poisson's equation. By this way the holographic principle   does't contradict `special theory of relativity' but it is likely  not to be valid in curved space time, at least in the case of strong gravity!

\section{Discussion} 
 
 Taken as a whole, our model and Verlinde's approach may be seen as playing complementary roles. In 
Verlinde's approach Newton's  universal law of gravitation is a consequence of certain thermal
and entropic properties of the constituents of spacetime, whereas in our model these properties
appear in a reverse consequence. First we consider that spinors (massive or massless) are everywhere in the space-time and  axial currents are proportional to killing vectors and then the gravitational potential $V$ is nothing but the logarithm of the proportionality factor. In static case together with weak field approximation  the development of $V$ is given by the Poisson's  equation of Newtonian Gravity. Then moving in reverse order to that of Verlinde's approach  we get holographic principle as a logical consequence such that Newton's constant plays the role of minimal unit of surface area.  It appears that, though not `action at a distance', the holographic principle is valid only in the weak field case!


\end{document}